\documentclass{article}
\usepackage{spconf,amsmath,graphicx}
\usepackage{amsfonts, amssymb, multirow}

\title{Audio Cross Verification Using Dual Alignment Likelihood Ratio Test}
\name{Heidi Lei$^{\dagger \star}$ \qquad Arm Wonghirundacha$^{\ddagger \star}$ \qquad Irmak Bukey$^{\ddagger \star}$ \qquad TJ Tsai$^\mathsection$\thanks{$^\star$Equal contribution}\thanks{\copyright~2023 IEEE. Personal use of this material is permitted. Permission from IEEE must be obtained for all other uses, in any current or future media, including reprinting/republishing this material for advertising or promotional purposes, creating new collective works, for resale or redistribution to servers or lists, or reuse of any copyrighted component of this work in other works.}}
\address{$^\dagger$Massachusetts Institute of Technology \\
	$^\ddagger$Pomona College \\
	$^\mathsection$Harvey Mudd College}
\begin{document}
\ninept
\maketitle
\begin{abstract}
This paper explores a way to verify that audio has not been maliciously tampered in a specific context: short viral videos taken from news recordings.  Rather than trying to detect artifacts of tampering (internal inconsistency), we focus on positively verifying a query against a trusted source such as a news recording (external consistency).  We propose a method for cross verifying a short audio query against a reference recording from which it was taken.  Our approach is to define two hypotheses (non-tampered vs tampered), calculate the most likely alignment between query and reference for each hypothesis, and then perform a likelihood ratio test on the two alignments.  We show that this method is fast to compute, much more robust than using MFCC features with Euclidean distance, and has the key benefit of explainability.  Our cross verification approach provides an alternative perspective and complementary tool to existing tampering detection methods.
\end{abstract}
\section{Introduction}
\label{sec:intro}

One significant issue in our society today is the reliability of audiovisual information.  The availability of deep fake technology and audio/video digital editing software has made it easy for non-experts to generate or modify audiovisual content in a way that seems realistic.  These technologies have been used for nefarious purposes such as nonconsensual pornography and political defamation \cite{westerlund2019emergence}, and their use casts doubt on the authenticity of legitimate audiovisual data \cite{vaccari2020deepfakes}.

This paper focuses on a very specific subset of this general problem: verifying the authenticity of viral videos posted on social media whose source content was directly recorded by major news agencies.  This includes speeches by political leaders or other high profile individuals, which tend to be on matters of national interest and thus are especially important to consider.  One could imagine a company like Twitter providing a green check mark next to a viral video that says ``Verified against NBC” (or some other major news agency), which would provide the viewer confidence that what they are watching can be trusted and has not been tampered in a malicious way.  On a technical level, this problem has two inputs: a short query (e.g. a low-res 10 second video clip of a campaign speech that is posted on social media) and a long reference recording from which the query is taken or adapted (e.g. a high-res 30 minute recording of the entire campaign speech as recorded by a major news agency).  The goal is then to compare the query and reference, and to determine if the query matches the reference (verified matching) or if it has been tampered (verified tampered).  In this work, we focus exclusively on verifying audio content.  This is the audio cross verification problem.

There is a large body of previous work on detecting audio tampering.  Audio tampering involves modifying a genuine audio recording through insertion, deletion, or replacement with foreign material.  Many audio forensics methods have been proposed to detect artifacts of tampering, such as double compression \cite{buker2021angular}\cite{luo2016detection}, broken watermarks \cite{nita2018tic}\cite{dobre2019tic}, discontinuities in the embedded electrical network frequency (ENF) \cite{wang2020enf}\cite{lin2017supervised}, inconsistencies in the acoustic environment signature \cite{bhangale2019tampering}\cite{meng2018detecting}, or singularities in the spectrogram \cite{zhang2022robust}\cite{chuchra2022deep}\cite{liu2021lightcvt}.  See \cite{bevinamarad2020audio} for a survey of previous work.  The timeliness of this topic is shown by the recent introduction of new datasets for speaker verification \cite{zhang2022partialspoof} and Deepfake video detection \cite{cai2022you} to study short duration temporal forgeries.  While many of the above methods work well in detecting tampering, they can also be extremely vulnerable to anti-forensic countermeasures, as recent studies have shown \cite{liu2022anti}\cite{tao2020anti}\cite{yan2019detection}\cite{li2019anti}.  

While similar in application, the audio cross verification problem has a different objective than audio tampering detection.  Rather than trying to detect artifacts of tampering (internal inconsistency), it instead focuses on positively verifying a query against a trusted source (external consistency).\footnote{Reference-based audio authentication is not new – it has been explored in the audio forensics community, primarily through extracting the ENF signal from the query and comparing against an ENF database from the electric grid.  This technique can be used to ascertain the timestamp of recording and assess if any tampering has been done (e.g. \cite{hua2018error}\cite{hua2016audio}\cite{lv2013audio}\cite{grigoras2007applications}).  In this paper, we adopt a similar approach but focus on directly verifying the audio content.}  This change in paradigm has two significant benefits and one significant drawback.  The first benefit is that the efficacy of a cross verification method does not depend on how seamlessly a tampering operation is done (since it is not looking for artifacts of tampering).  In this way, it provides a permanent and stable solution rather than one that depends on the state of deepfake technology or anti-forensic countermeasures.  The second benefit is that cross verification methods can make a stronger claim than audio tampering methods.  An audio tampering method can at best say, ``There is no evidence of tampering (though it’s possible that the tampering was done seamlessly or the whole recording is a deepfake)”.  A cross verification method can say, ``This clip matches a trusted source, so it can be trusted.”  The drawback of cross verification methods is that they can only be used when a trusted reference is available, which limits the contexts in which they can be used.  

An ideal audio cross verification method would have several desirable characteristics.  First, it should be extremely accurate in predicting whether a query is matching or tampered.  Second, it should be robust to real-world conditions.  In particular, we would want the method to be insensitive to harmless differences like audio encoding formats, volume differences, and audio compression artifacts, while at the same time being very sensitive to changes due to tampering operations like insertion, deletion, and replacement.  We would also want the method to be robust to overfitting, since complex models can be vulnerable to distribution shift.  Third, an ideal method would be fast in terms of runtime.  Fourth, the method should be explainable.  Since this tool would be public facing and have ramifications in the public sphere, it is necessary for predictions to be justifiable.

We propose an approach called Dual Alignment Likelihood Ratio Test (DA-LRT) that is designed around the desirable characteristics above.  DA-LRT consists of three stages.  The first stage is pre-selection, in which we use a binary feature representation to quickly determine the rough offset where the query occurs in the reference.  The second stage is to perform two different alignments between the query and the reference.  One alignment is performed under the hypothesis that there is no tampering, and the other alignment is performed under the hypothesis that tampering is present in the form of insertions, deletions, and replacements.  The third stage is to identify differences between the two alignments, and then to perform a statistical test to determine which hypothesis is more likely given the observed features.  We describe this in detail in the next section.


\section{System Description}
\label{sec:systemDescr}

\begin{figure}
	\includegraphics[width=\linewidth]{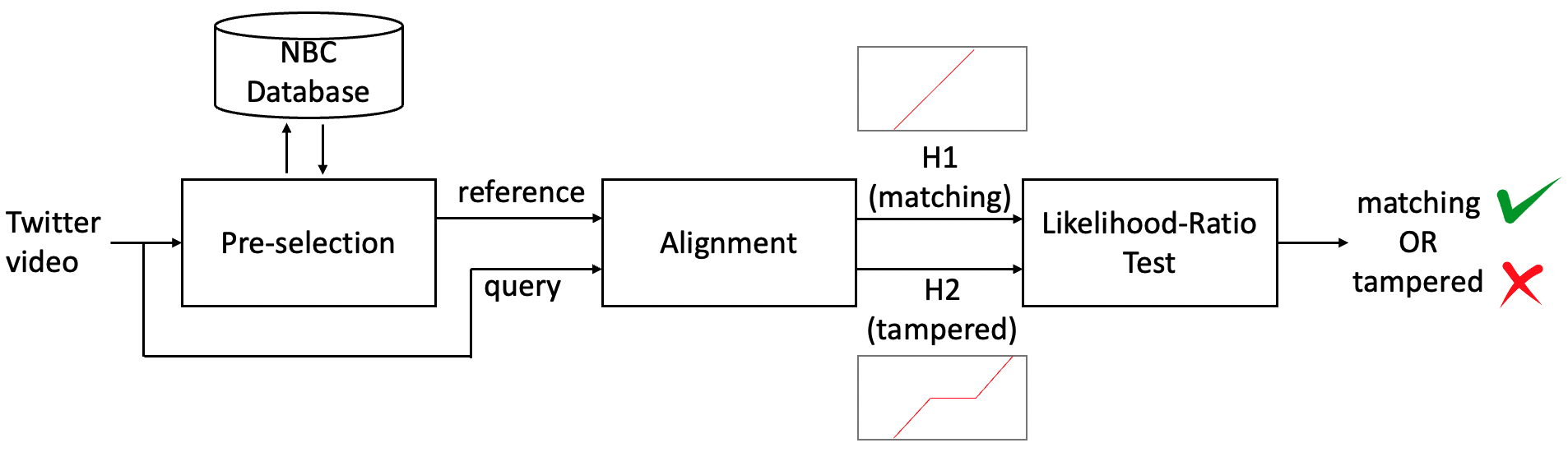}
	\caption{Overview of the proposed Dual Alignment Likelihood Ratio Test (DA-LRT) method for audio cross verification.}
	\label{fig:systemOverview}
\end{figure}

The DA-LRT method has three main stages (see Figure \ref{fig:systemOverview}), which are described in detail in the next three subsections.\footnote{Code at https://github.com/HMC-MIR/AudioCrossVerification.}  Our goal is to compare a short query recording to a reference recording (from which it was taken), and to determine if the query matches the reference or if it has been tampered.  

\subsection{Pre-Selection}
\label{subsec:preselection}

The first stage is pre-selection, in which the goal is to find the original source material that matches the query.  File-level pre-selection can be done using an audio fingerprinting method (e.g. Shazam \cite{wang2003industrial}, Google \cite{baluja2007audio}) or metadata search (e.g.~speaker and date).  To keep our discussion focused, we will simply assume in this paper that a matching reference recording has already been identified.\footnote{Note that, if the query is mostly tampered material or is a completely synthetic deep fake, a match will not be found in the database.  In our Twitter application scenario, one could affix a label that says ``No match found with NBC" to provide useful information to the viewer.}  

Given a matching reference recording, we would also like to perform temporal pre-selection in order to identify the part of the reference recording that matches the query.  Because stage 2 requires dynamic programming, this temporal pre-selection can significantly reduce total runtime if the reference is much longer than the query.  We use a simplified version of audio hashprints \cite{tsai2017known} to accomplish this quickly and efficiently.  This approach consists of three steps.  First, we compute standard mel frequency cepstral coefficients (MFCCs) with 25 ms analysis frames and 10 ms hop size.  We include delta and delta-delta features, which results in a total of 39 features per frame.  Second, we represent each frame with a 26-bit binary feature representation by thresholding the delta and delta-delta features at 0.  (The MFCC features are stored for later use in stages 2 and 3.)  Third, we identify the offset in the reference recording that results in the lowest total Hamming distance between the corresponding binary feature sequences (assuming a 1-to-1 correspondence).  This can be computed efficiently by encoding each frame in memory as a single 32-bit integer and performing bit operations.  The optimal offset specifies the approximate matching region of the reference recording.  Since the query may be tampered from insertion or deletion, we include a short buffer before and after the matching region to provide a conservative estimate.  This specifies our pre-selected reference material.

\subsection{Dual Alignment}
\label{subsec:alignment}

The second stage is to compute two different alignments between the query and the pre-selected reference material.  These two alignments are estimated under two different hypotheses (tampered vs.~non-tampered).  The procedure for the two alignments is described in detail in the remainder of this subsection.  We will refer to the query features as $q_0$, $q_1$, \dots, $q_{N-1}$ and the (pre-selected) reference features as $r_0$, $r_1$, \dots, $r_{M-1}$, where $q_i \in \mathbb{R}^{39}$ and $r_i \in \mathbb{R}^{39}$ are the MFCC features for the $i^{th}$ audio frame.

The first alignment is estimated under the assumption that the query is non-tampered.  In this case, the alignment path is assumed to be a diagonal line, and the only unknown is the offset at which the matching region begins.  We re-estimate\footnote{We re-estimate the offset using MFCCs since they are more informative than coarsely quantized binary hashprints in the pre-selection stage.} the alignment path by (1) computing a pairwise cost matrix $C \in \mathbb{R}^{N \times M}$ between the query and reference using the MFCC features and a Euclidean distance cost metric, (2) calculating the sum of pairwise costs along each complete diagonal path of the cost matrix, and (3) selecting the diagonal alignment path with the minimum total cost.  The optimal path has (query, reference) coordinates (0,$\Delta$), (1, $\Delta+1$), \dots, ($N-1$, $\Delta+N-1$).

The second alignment is estimated under the assumption that the query is tampered.  In this case, the shape of the alignment path is not known in advance, since the query may have been tampered with insertions, deletions, and replacements.  We estimate this alignment path using an adaptation of Hidden State Time Warping (HSTW) \cite{chang2022parameter}, a previously proposed dynamic programming algorithm that allows for state-based time warping.  HSTW allows the alignment path at any location in the cost matrix to be in one of two states, where each state has its own unique time warping characteristics.  In our scenario, we define one state to be a ``matching" state in which the only allowable transition is (1,1), and we define the other state to be a ``tampered" state in which the allowable transitions are (0,1) and (1,0).  For completeness, we describe the algorithm below.

Our adaptation of HSTW has four steps.  First, we calculate a pairwise cost matrix $C \in \mathbb{R}^{N \times M}$ using MFCCs and Euclidean distance.  Second, we initialize the cumulative cost tensor $D \in \mathbb{R}^{2 \times N \times M}$ and corresponding backtrace tensor $B \in \mathbb{N}^{2 \times N \times M}$.  Note that the two ``planes" of the tensor correspond to the two different states.  We will refer to the tampered plane as $D_t \in \mathbb{R}^{N \times M}$ and the matching plane as $D_m \in \mathbb{R}^{N \times M}$.  Since the query occurs at an unknown offset in the reference, we initialize $D_m[0,j] = C[0,j]$, $j=0,1, \dots, M-1$ and $D_t[0,j] = \frac{\alpha + \gamma}{2}$, $j=0,1, \dots, M-1$, where $\alpha$ and $\gamma$ are hyperparameters (described in more detail below).  This initialization allows the alignment path to begin at any offset in the reference in either state without penalty.  Third, we compute the remaining values in $D[i,j]$, $i=1,\dots,N-1$, $j=0,\dots,M-1$ using the following dynamic programming rules: $D_t[i,j] = min(D_t[i,j-1] + \gamma, D_t[i-1,j] + \alpha, D_m[i-1,j-1] + \gamma + \alpha)$ and $D_m[i,j] = min(D_m[i-1,j-1] + C[i,j], D_t[i,j] + \beta)$.  Here, $\alpha$, $\gamma$, and $\beta$ are hyperparameters specifying the insertion, deletion, and plane transition penalties, respectively.  As we compute each element in $D$, we also update the corresponding element of $B$ to record the optimal transition type.  Fourth, we identify the element in the last row of $D_t$ or $D_m$ that has the lowest cumulative cost, and then follow the backpointers in $B$ to determine the optimal alignment path.

\subsection{Likelihood-Ratio Test}
\label{subsec:likelihoodTest}

The third stage is to treat each alignment path as a hypothesis, and then to determine which hypothesis has a higher likelihood based on the observed features.  This stage consists of four steps, which are described below.  For brevity, we will refer to the matching hypothesis as H1 and the non-matching hypothesis as H2.  

The first step is to identify differences between the alignment paths under H1 and H2.  Specifically, we partition the query frames $q_0$, $q_1$, \dots, $q_{N-1}$ into two groups: those whose alignments are the same between H1 and H2 (partition $A$), and those whose alignments are different (partition $B$).  To account for analysis frame offsets, we define ``different" to mean that the alignments differ by 2 or more frames.  Query frames whose HSTW alignment lies entirely in the tampering plane are placed in partition $B$.  If the two alignment paths are identical, we simply declare the query to be matching.

The second step is to model the observed differences between query and reference features in matching regions (i.e.~where the query is non-tampered).  We do not know in advance which frames are matching, but we can interpret regions where the two alignment paths agree as circumstantial evidence that such regions are matching.  Accordingly, we use the observed features for query frames in partition A to model what a ``match" looks like.  Let ($n_1$, $m_1$), ($n_2$, $m_2$), \dots, ($n_L$, $m_L$) be the alignment path coordinates where the two alignment paths are in agreement, $q_{n_i} \in \mathbb{R}^{39}$ and $r_{m_i} \in \mathbb{R}^{39}, i=1,\dots,L$ denote the corresponding MFCC feature vectors, and $q_{n_i, f} \in \mathbb{R}$ and $r_{m_i, f} \in \mathbb{R}$ indicate the $f^{th}$ MFCC feature in $q_{n_i}$ and $r_{m_i}$, respectively.  For each of the 39 features, we calculate the mean and variance of the observed feature differences in matching regions as $\mu_f = \frac{1}{L} \sum_{i=1}^{L} (q_{n_i, f} - r_{m_i, f} )$ and $\sigma_f^2= \frac{1}{L-1} \sum_{i=1}^{L} \Big( (q_{n_i, f} - r_{m_i, f}) - \mu_f \Big)^2$.  The end result is a set of 39 scalar Gaussian distributions $\mathcal{N}(\mu_f, \sigma_f^2)$, $f=1,\dots,39$ that model the feature differences when the query and reference frames are matching.  We will use these to compute the likelihood of H1.

The third step is to model the differences between query and reference features in non-matching regions (i.e.~where the query is tampered).  In tampered regions, we assume that the query and reference features are independent.  If we treat each feature as an independent random variable drawn from a scalar Gaussian distribution $\mathcal{N}(\mu_f, \sigma_f^2)$ (as estimated above), then the mean and variance of feature differences in non-matching regions can be estimated as $\tilde{\mu}_f = \frac{1}{L} \sum_{i=1}^{L} (q_{n_i, f} - r_{m_i, f} )$ and $\tilde{\sigma}_f^2 = \frac{1}{L-1} \sum_{i=1}^{L} \Big[ (q_{n_i, f} - \mu_f^{query})^2 + (r_{m_i, f} - \mu_f^{ref})^2 \Big]$, $f=1,\dots,39$.  The end result is again a set of 39 scalar Gaussian distributions $\mathcal{N}(\tilde{\mu}_f, \tilde{\sigma}_f^2)$, $f=1, \dots,39$ that model the feature differences in non-matching regions.  We will use these to compute the likelihood of H2.

The fourth step is to perform a likelihood-ratio test on the two hypotheses using the query frames in partition $B$ as observations.  In other words, we would like to know which hypothesis is more likely given the observed feature differences.  To make the solution tractable, we assume that features are independent across time and across feature type.  Clearly, these assumptions are faulty (e.g.~the features are highly correlated over time), but we apply these simplifying assumptions to both hypotheses, so we don’t anticipate that they will cause a systematic bias in one direction.  We describe the details of the likelihood-ratio test in the next two paragraphs.

\textit{Definitions}.  Let the query frame indices in partition $B$ be given by $\tilde{n}_1$, $\tilde{n}_2$, \dots, $\tilde{n}_{N-L}$.  (There are $L$ query frames in partition $A$, so partition $B$ will have $N-L$ elements.)  Let the alignment path coordinates for these query frames under H1 be given by ($\tilde{n}_1$, $\Delta + \tilde{n}_1$), ($\tilde{n}_2$, $\Delta + \tilde{n}_2$), \dots, ($\tilde{n}_{N-L}$, $\Delta + \tilde{n}_{N-L}$) and the alignment path coordinates under H2 be given by ($\tilde{n}_1$, $\tilde{m}_1$), ($\tilde{n}_2$, $\tilde{m}_2$), \dots, ($\tilde{n}_{N-L}$, $\tilde{m}_{N-L}$).  The feature differences are modeled by $\mathcal{N}(\mu_f, \sigma_f^2)$, $f=1,\dots,39$ under H1 and by $\mathcal{N}(\tilde{\mu}_f, \tilde{\sigma}_f^2)$, $f=1,\dots,39$ under H2.

\textit{Likelihood Calculation}.  Given the independence assumptions, the maximum likelihood criterion is given by:
\begin{multline}
\label{eqn:mlcriterion}
	\prod_{i=1}^{N-L} \prod_{f=1}^{39} \frac{1}{\sqrt{2\pi \sigma_f^2}} e^{-\frac{1}{2 \sigma_f^2} \big( (q_{\tilde{n}_i, f} - r_{\Delta + \tilde{n}_i, f}) - \mu_f \big)^2} 
	\underset{H2}{\overset{H1}{\gtrless}}  \\
	\prod_{i=1}^{N-L} \prod_{f=1}^{39} \frac{1}{\sqrt{2\pi \tilde{\sigma}_f^2}} e^{-\frac{1}{2 \tilde{\sigma}_f^2} \big( (q_{\tilde{n}_i, f} - r_{\tilde{m}_i, f}) - \tilde{\mu}_f \big)^2}
\end{multline}
Note that the two sides of equation \ref{eqn:mlcriterion} use different observations (i.e.~$(q_{\tilde{n}_i, f} - r_{\Delta+\tilde{n}_i, f})$ vs $(q_{\tilde{n}_i, f} - r_{\tilde{m}_i, f})$), whereas a typical maximum likelihood formulation has a shared set of observations.  Taking the log of both sides, equation \ref{eqn:mlcriterion} simplifies to:
\begin{equation}
\label{eqn:mlsimplified}
	\sum_{i=1}^{N-L} \sum_{f=1}^{39} \Big( ln(\frac{\tilde{\sigma}_f^2}{\sigma_f^2}) - z_{i,f}^2  + \tilde{z}_{i,f}^2 \Big)
	\underset{H2}{\overset{H1}{\gtrless}} 0 \\
\end{equation}
where $z_{i,f} \triangleq \frac{(q_{\tilde{n}_i, f} - r_{\Delta + \tilde{n}_i, f}) - \mu_f}{\sigma_f}$ and $\tilde{z}_{i,f} \triangleq \frac{(q_{\tilde{n}_i, f} - r_{\tilde{m}_i, f}) - \tilde{\mu}_f}{\tilde{\sigma}_f}$ are the Z-scores of the observed feature differences under H1 and H2, respectively.  The left side of equation \ref{eqn:mlsimplified} is a log likelihood-ratio test statistic, and we use this quantity as the final ``tampering score" for the query.  This allows us to characterize performance more broadly with a receiver operation characteristic curve.

\section{Experimental Setup}
\label{sec:expSetup}

We used the DAPS dataset \cite{mysore2014can} for our experiments.  This dataset contains high-quality audio recordings of 20 different speakers each reading 5 scripts, where scripts are 2--4 minutes long.  The dataset also includes recordings in multiple acoustic conditions, which will allow us to study the effect of acoustic environment in future work.  For this study, we focus only on the 100 high-quality, low-noise recordings since these most closely match our intended application.

The queries are generated by processing the DAPS data in the following way.  First, each of the 100 high quality original recordings is compressed to a lower bitrate at R kbps.  This step reflects the fact that videos shared on social media are typically compressed.  Second, we randomly sample 10 different 10-second segments from each compressed recording.  Third, we tamper each 10-second segment in three different ways to simulate insertion, deletion, and replacement tampering operations.  For insertions, we randomly select an L second filler segment from within the same compressed recording (but outside of the selected 10-second segment).  This L second filler is then inserted into the segment at a random location in the segment.  This arrangement means that the inserted foreign material is perfectly matched in terms of speaker, speaking style, and acoustic environment.  For deletions, we randomly select an L second interval from within the 10-second segment and delete it.  For replacements, we randomly select an L second interval from within the 10-second segment, and then replace it with another L second filler randomly selected from within the same compressed recording.  In addition to the tampered versions, we also include an untampered version.

Our benchmarks are constructed in the following manner.  Given the above method for generating queries, we have a total of 100 x 10 x 4 = 4000 queries for a given bitrate R and tampering duration L.  Each of these queries is associated with a reference recording, which is the original DAPS recording from which the query was taken.  Half of the speakers are set aside for training, and the other half are set aside for testing.  We consider three different bitrates R = 256, 128, and 64 kbps and five different tampering durations L = 4, 2, 1, 0.5, 0.25 seconds.  In total, our evaluation includes 15 benchmarks, each containing 2000 training queries and 2000 test queries.

Our evaluation metric is equal error rate (EER).  EER is a useful metric because it summarizes performance in a single number and is invariant to priors.  Note that we chose to evaluate classification performance at the recording (rather than frame) level since this most closely aligns with the user’s experience of our intended application.

\section{Results}
\label{sec:results}

Table 1 shows the results of our proposed method on the audio cross verification task.  The left half of Table \ref{tab:results} shows the performance of the proposed system, and the right half shows the performance of a baseline system (``MFCC-Euclidean") that uses the Euclidean distance between MFCC features as a tampering score.  The table shows the EER of both systems across a range of different conditions.  (Note that numbers are expressed in percentages, so 0.20 corresponds to a 0.20\% EER.)  Each row shows the performance with a fixed tampering duration $L$ and query bitrate $R$, and each column shows the performance in detecting tampering of specific types (i.e.~the benchmark only includes one tampering type plus untampered queries).  The rightmost column (``all") in each panel shows performance when all tampering types are present, and the bottommost row (``all") in each panel shows performance when all tampering durations are present.  

\begin{table}
	\begin{center}
		\begin{tabular}{|c|c|cccc|cccc|} 
			\hline
			& Tamp & \multicolumn{4}{|c|}{DA-LRT} & \multicolumn{4}{|c|}{MFCC-Euclidean} \\
			& Len & ins & del & rep & all & ins & del & rep & all \\			
			\hline
			\parbox[t]{1mm}{\multirow{5}{*}{\rotatebox[origin=c]{90}{256 kbps}}}
			& 4.0s & .00 & .00 & .00 & .00 & .00 & 0.4 & .00 & .13  \\
			& 2.0s & .00 & .00 & .00 & .00 & .20 & 1.6 & 1.2 & 1.0 \\
			& 1.0s & .00 & .00 & .00 & .00 & 1.6 & 2.0 & 12.0 & 6.2 \\
			& 0.5s & .20 & .20 & 3.1 & 1.2 & 4.6 & 2.2 & 29 & 15 \\
			& 0.25s & .20 & .00 & 33 & 14 & 5.4 & 3.2 & 40 & 20 \\ 
			& all & .08 & .04 & 9.5 & 3.4 & 2.7 & 2.0 & 20 & 9.2 \\ 
			\hline
			\parbox[t]{1mm}{\multirow{5}{*}{\rotatebox[origin=c]{90}{128 kbps}}}
			& 4.0s & .00 & .00 & .00 & .00 & .00 & .40 & .00 & .13  \\
			& 2.0s & .00 & .00 & .00 & .00 & .20 & 1.6 & 1.0 & .93 \\
			& 1.0s & .00 & .00 & .00 & .00 & 1.4 & 1.8 & 12 & 5.9 \\
			& 0.5s & .20 & .40 & 3.1 & 1.3 & 4.6 & 2.0 & 30 & 15 \\
			& 0.25s & .20 & .00 & 34 & 15 & 5.4 & 3.0 & 40 & 20 \\ 
			& all & .08 & .08 & 10 & 3.7 & 2.6 & 1.9 & 20 & 9.0 \\ 
			\hline
			\parbox[t]{1mm}{\multirow{5}{*}{\rotatebox[origin=c]{90}{64 kbps}}}
			& 4.0s & .00 & .00 & .00 & .00 & .00 & .20 & .00 & .07  \\
			& 2.0s & .00 & .00 & .00 & .00 & .20 & 1.0 & 0.4 & .53 \\
			& 1.0s & .00 & .00 & .20 & .07 & 1.0 & 1.8 & 10 & 5.1 \\
			& 0.5s & .20 & .40 & 4.0 & 1.6 & 3.8 & 1.8 & 28 & 14 \\
			& 0.25s & .20 & .00 & 38 & 17 & 4.8 & 2.6 & 39 & 19 \\ 
			& all & .08 & .08 & 12 & 4.3 & 2.2 & 1.5 & 19 & 8.8 \\ 
			\hline
		\end{tabular}
	\end{center}
	\caption{Comparing the performance of DA-LRT and MFCC-Euclidean distance baseline on the audio cross verification task.  Numbers indicate equal error rate in percentages, so 0.20 corresponds to 0.20\% EER.  Rows show performance for a fixed tampering duration and query bitrate, and columns show performance in detecting specific types of tampering (insertions, deletions, replacements).  The rightmost column and bottommost row in each panel show aggregate performance when multiple tampering types and multiple tampering durations are present, respectively.}
	\label{tab:results}
\end{table}

There are three things to notice about the results in Table \ref{tab:results}.  First, queries tampered with replacement have much higher error rates than queries tampered with insertion or deletion.  This is because insertion and deletion operations will cause any audio frames after the tampering point to become unsynchronized with the reference, making it much easier to detect differences.  In contrast, tampering through replacement does not cause this global shift, so it can only be detected by observing feature differences within the actual tampered region.  It should be noted that for short tampering operations (e.g. L=0.25), our method of randomly selecting intervals may result in replacing silence with silence, so these results may be overly pessimistic.  Second, the proposed method substantially reduces the duration of tampering that can be detected reliably.  We see that the MFCC-Euclidean baseline has low (but not perfect) error rates for tampering durations of 2 seconds or more, whereas the proposed method has reliable performance for insertion and deletion queries down to L=0.25 sec and for replacement queries down to 0.5-1.0 sec.  Third, query bitrate does not affect performance with long tampering durations, but has a moderate effect with short tampering durations.  For example, the EER for replacement queries with L=0.5 sec is 3.1\% for R=256 and R=128, but worsens to 4.0\% for R=64.  Because 64 kbps is considered a very low bitrate, we can interpret these results as a kind of worst case scenario.

It is useful to consider the strengths and weaknesses of our proposed method.  Recall the four characteristics that we identified in an ideal solution to the cross verification problem: accurate, robust, fast, and explainable.  With regards to accuracy and robustness to real-world conditions, our proposed method reliably detects insertions and deletions, since these cause a global shift that results in feature mismatches for all audio frames after the tampering point.  The biggest weakness of the system is in detecting short duration ($<0.5$ sec) replacements, which must be detected solely on local feature mismatches.  With regards to having a fast runtime, the system is fast enough to be useful in an automated system.  In our experiments, it took an average of 433 ms to compute the MFCC features on the query, 5.82 sec to compute the MFCC features on the reference, and 123 ms to perform the remaining computations for offset estimation, alignments, and maximum likelihood test.\footnote{We used the python\_speech\_features library for MFCC feature computation, implemented the alignment in cython, and ran our experiments on a 2.4 GHz Intel Xeon CPU.}  With regards to explainability, our method compares the likelihood of two interpretable hypotheses (tampered vs non-tampered) and makes clear, explicit assumptions about the nature of the alignment and likelihood calculations.

\section{Acknowledgments}
This material is based upon work supported by the National Science Foundation under Grant No.~1948531.

\bibliographystyle{IEEEbib}
\bibliography{AudioCrossVerification}

\end{document}